\documentclass[11pt]{article}
\usepackage[left=0.6in,right=0.6in]{geometry}
\usepackage{empheq}
\usepackage{amssymb}
\usepackage{dsfont}
\usepackage{slashed}
\usepackage[pdfstartview=FitH,pdfpagemode=None]{hyperref}
\usepackage{color}
\usepackage{multirow}
\usepackage{mathtools}
\usepackage{stackrel}

\newcommand{\nn}{\nonumber}
\newcommand{\be}{\begin{equation}}
\newcommand{\ee}{\end{equation}}
\newcommand{\ba}{\begin{eqnarray}}
\newcommand{\ea}{\end{eqnarray}}
\newcommand{\bea}{\begin{eqnarray}}
\newcommand{\eea}{\end{eqnarray}}

\newcommand{\vev}[1]{{\left< {#1} \right>}}

\begin{document}

\begin{titlepage}
\hfill LCTP-18-05

\begin{center}

{\Large \textbf{Zeta-function Regularization of Holographic Wilson Loops}}\\[4em]

{\small  Jerem\'ias Aguilera-Damia${}^1$,  Alberto Faraggi$^{2}$, Leopoldo A.~Pando Zayas${}^{3,4}$, Vimal Rathee${}^{3}$ and Guillermo A. Silva${}^{1,4}$}\\[3em]
%
%

%{\large }
${}^{1}$\emph{Instituto de F\'isica de La Plata - CONICET \& Departamento de F\'isica\\
 UNLP C.C. 67, 1900 La Plata, Argentina}\\[1em]

${}^{2}$\emph{Departamento de Ciencias F\'isicas,  Facultad de Ciencias Exactas\\
Universidad Andr\'es Bello,  Sazie 2212, Piso 7, Santiago, Chile}\\[1em]

${}^{3}$\emph{Leinweber Center for Theoretical Physics,  Randall Laboratory of Physics\\ The University of
Michigan,  Ann Arbor, MI 48109, USA}\\[1em]

${}^4$\emph{The Abdus Salam International Centre for Theoretical Physics\\ Strada Costiera 11,  34014 Trieste, Italy}\\[6em]

\abstract{Using $\zeta$-function regularization, we study the one-loop effective action of fundamental strings in $AdS_5\times S^5$ dual to the latitude  $\frac{1}{4}$-BPS Wilson loop in $\mathcal{N}=4$ Super-Yang-Mills theory.  To avoid certain ambiguities inherent to string theory on curved backgrounds we subtract the effective action of the holographic  $\frac{1}{2}$-BPS Wilson loop. We find agreement with the expected field theory result at first order in the small latitude angle expansion but discrepancies at higher order.}

\vspace{2cm}
{\tt jeremiasadlp@gmail.com, alberto.faraggi@unab.cl, lpandoz@umich.edu,\\ vimalr@umich.edu, silva@fisica.unlp.edu.ar}
\end{center}

%\paragraph{Keywords:} 
\end{titlepage}

\newpage

\section{Introduction}
One of the most powerful promises embodied by the  AdS/CFT correspondence is to broaden our perspective of string theory beyond our limited understanding in flat  spacetimes.  We expect that some of the main principles, such as conformal invariance, retain its central place. However, most of the computational technology available for strings in flat spacetimes is inadequate in the more general setup of strings in curved spaces. 

Thus far many questions in AdS/CFT  at the quantum level have been dealt with on a case by case basis without a  general framework. With the advent of localization techniques the panorama has changed radically as the exact answer to many problems are now known.  This opens the possibility of testing different computational methods. The study of quantum corrections in the case of Wilson loops is particularly promising in the context of the AdS/CFT correspondence because the expectation value of Wilson loops is determined by string world-sheets \cite{Maldacena:1998im,Rey:1998ik} and consequently pushes us to confront the underpinnings of string perturbation theory more directly. 

Some of the most studied examples at the quantum level are the holographic duals of the $\frac{1}{2}$-BPS and $\frac{1}{4}$-BPS Wilson loops in $\mathcal{N}=4$ SYM.  In the semi-classical approximation the one-loop corrections are equivalent to computations of determinants of certain Laplace-like operators in curved spaces.  Determinant of operators in curved space have a long history in physics and  also in mathematics as sources of spectral information. There are, indeed, various computational methods that have already been applied in the context of holographic Wilson loops.  For example, the expectation value of the holographic  $\frac{1}{2}$-BPS  Wilson loop was originally computed using $\zeta$-function techniques in \cite{Drukker:2000ep}  and subsequently revisited using the Gelfand-Yaglom approach in \cite{Kruczenski:2008zk}.  More recently the better-defined  problem of computing the difference of the effective actions of the holographic $\frac{1}{4}$- and  $\frac{1}{2}$-BPS strings  has received particular attention since supersymmetric localization provides a precise answer. The first attempts were reported in  \cite{Forini:2015bgo,Faraggi:2016ekd}. These two groups used a Gelfand-Yaglom based method to tackle the problem but did not find a match with the field theory prediction. Ultimately, after a careful analysis, the mismatch was traced back to a change in topology from the disk to the cylinder and the use of a diffeomorphism-invariant IR cutoff  \cite{Cagnazzo:2017sny}. 
%{\color{red} In an interesting recent development reported in \cite{Forini:2017whz}, a perturbative computation on the disk using heat kernel methods was shown to match the field theory answer  to first order in the latitude angle. La sacamos? aparece de vuelta la misma oracion mas abajo al ppio del parrafo}

A more immediate motivation for developing $\zeta$-function regularization techniques stems from the fact that using perturbative heat kernel techniques to the first nontrivial order in the latitude angle, the authors  of \cite{Forini:2017whz} found a match between the gauge and gravity calculations for the expectation value of the $\frac{1}{4}$-BPS latitude Wilson loop. This suggests that  $\zeta$-function might be the correct framework to compute the one-loop determinants for the spectrum of fluctuations of the string; it also attacks the problem directly on the disk rather than mapping it to the cylinder as done in  \cite{Kruczenski:2008zk,Forini:2015bgo,Faraggi:2016ekd,Cagnazzo:2017sny}.  The holographic  dual to the  $\frac{1}{2}$-BPS  Wilson loop is a fundamental string with  $AdS_2$ worldsheet. For this homogeneous space one can address its one-loop effective action with results dating back to \cite{Camporesi:1994ga,Camporesi:1995fb} as was done in \cite{Drukker:2000ep,Buchbinder:2014nia}. For the  $\frac{1}{4}$-BPS,  however, the space is no longer homogeneous and new technology is required to evaluate the determinants.  In this manuscript we approach the computation of one-loop determinants using recent results of  $\zeta$-function regularization of Laplace-like operators in conformally $AdS_2$ spaces  that are reported in a separate publication \cite{ZetaPaper}. There is a strong general motivation to develop $\zeta$-function regularization. Starting with the insightful works of \cite{Dowker:1975tf},\cite{Hawking:1976ja}, $\zeta$-function regularization methods have shown to be highly reliable in various areas of applications \cite{Elizalde:1994gf}; we hope that generalizing such methods will find natural applications in several contexts.  

We show that the $\zeta$-function regularized answer matches at leading order in the small latitude angle but receives correction at higher order, leading to a mismatch with the expected field theory answer.

The rest of the paper is organized as follows. In section \ref{Sec:latitude} we briefly review some of the most salient features of the semiclassical approach to holographic Wilson loops. Section \ref{sec: results} presents a summary of the result of our companion paper  where we obtained explicit expressions for determinants of general Laplace-like operators in conformally $AdS_2$ spacetimes.  Section \ref{sec: Wilson loops} determines the ratio of the latitude to  the $\frac{1}{2}$-BPS holographic Wilson loops.  We conclude in section \ref{Sec:Conclusions}.

%%%%%%%%%%%%%%%%%%%%%%%%%%%%%%%%%%%%%
\section{Latitude Wilson loops}\label{Sec:latitude}

In order to make this paper somehow self-contained we briefly review some of the most salient features of the holographic Wilson loops we discuss. This subject has been the center of a lot of investigation recently and we refer the reader to the works \cite{Forini:2015bgo,Faraggi:2016ekd} for omitted details.

The  $\frac{1}{4}$-BPS  latitude Wilson loop (in the fundamental representation of $SU(N)$) is defined as \cite{Drukker:2006ga,Drukker:2007qr}
\begin{empheq}{align}
    W(C)=\frac{1}{N}\textrm{Tr}\,{\cal P}\exp\oint_C ds\left(iA_{\mu}\dot{x}^{\mu}+|\dot x|\Phi_I \, n^I(s)\right)\,,
\end{empheq}
where ${\cal P}$ denotes path ordering along the loop and $C$ labels a curve parametrized as
\bea
x^\mu(s)=(\cos s,  \sin s, 0,0)\,,\qquad
n^I(s)=(\sin\theta_0\cos s, \sin\theta_0\sin s, \cos\theta_0, 0,0,0),\qquad s\in(0,2\pi)
\eea
For $\theta_0=0$, this operator was shown to preserve half of the supersymmetries and its expectation value was evaluated exactly, under certain conjectures [Gaussian], by  \cite{Erickson:2000af} and \cite{Drukker:2000rr}. The definitive proof   was provided by Pestun via the by now thoroughly exploited supersymmetric localization technique \cite{Pestun:2007rz}. The answer, exact in the gauge group rank $N$ and the t' Hooft coupling $\lambda$, is
\bea
    \langle W\rangle_{\textrm{circle}}&=&\frac{1}{N}L^1_{N-1}\left(-\frac{\lambda}{4N}\right)e^{\lambda/8N}.       \label{halfBPSoperator}
\eea
More generally, for arbitrary values of $\theta_0$, the vacuum expectation value of this operator is conjectured to be given by a simple re-scaling of the 't Hooft coupling $\lambda\to\lambda'=\lambda \cos^2\theta_0$ in the above exact expression \cite{Drukker:2006ga,Drukker:2007qr,Drukker:2007yx}.

The dual $\frac{1}{2}$-BPS string has an $AdS_2\subset AdS_5$ worlsheet with disk topology, 
\be
\label{Eq:AdS2}
 ds^2 = d\rho^2+\sinh^2\!\rho\,\, d\tau^2, \quad
	\rho \geq0\,,  \quad 	\tau \sim\tau+2\pi\, .
	\ee
On the other hand, the $\frac{1}{4}$-BPS string worldsheet forms a cap through the north pole of  $S^2\subset S^5$ and the  induced geometry  is asymptotic to $AdS_2$,
\begin{empheq}{align}
   \label{induced}
   ds^2_M&=M (\rho)\,ds^2,   \qquad  M(\rho)=1+\frac{\sin^2\theta(\rho)}{\sinh^2\rho}\,, \qquad \sin\theta(\rho)=\frac{\sinh\rho\sin\theta_0}{\cosh\rho+\cos\theta_0}\,,
\end{empheq}
where $0\leq\theta_0\leq\frac{\pi}{2}$ is the latitude angle. The $\frac{1}{2}$-BPS solution corresponds to $\theta_0=0$. 
The string action can be evaluated on-shell on this classical solution. The result, after an appropriate renormalization, is  \cite{Drukker:2006ga}
\bea
S^{(0)}=-\sqrt{\lambda}\cos\theta_0\,.
\eea
Since $\langle W \rangle\simeq  \exp\left(-S^{(0)}\right)=\exp\left(\sqrt{\lambda}\cos\theta_0\right)$, we recover, at the leading classical level, the expectation (\ref{prediction}) from field theory.  

Comparing the one-loop effective actions of the  $\frac{1}{4}$ and $\frac{1}{2}$-BPS strings, as discussed in \cite{Forini:2015bgo,Faraggi:2016ekd}, and anticipated in \cite{Kruczenski:2008zk} leads to a better defined string theory problem since  both dual strings have world-sheets with disk topology. The general expectation is that the issues related to ghost zero modes and other aspects of string perturbation theory on curved spacetimes might cancel upon considering the difference of effective actions. The exact field theory answer at large $\lambda$ is 
\be
\frac{\vev{W}_\textrm{latitude}}{\vev{W}_\textrm{circle}}
\simeq\exp\left(
\sqrt{\lambda}(\cos\theta_0-1)-\frac{3}{2}\ln\cos\theta_0+\ldots
\right)\,.
\label{prediction}
\ee
The leading order term in the large $\lambda$ expansion was matched against a particular string worldsheet identified in \cite{Drukker:2006ga}. In recent years, there has been a strong effort in computing  the $-(3/2) \ln\cos\theta_0$ term from the string theory one-loop effective action \cite{Forini:2015bgo,Faraggi:2016ekd,Forini:2017whz,Cagnazzo:2017sny}. In this manuscript we approach this question using $\zeta$-function regularization.

At the semiclassical level, the fluctuations of the fundamental string dual to the $\frac{1}{4}$-BPS Wilson loop were thoroughly studied in \cite{Forini:2015bgo,Faraggi:2016ekd}. The spectrum involves the operators
\begin{empheq}{alignat=7}\label{eq: 1/4 BPS operators}
	\mathcal{O}_1(\theta_0)&=M^{-1}\left(-g^{\mu\nu}\nabla_{\mu}\nabla_{\nu}+2\right)\,,
	\qquad
	\mathcal{O}_2(\theta_0)=M^{-1}\left(-g^{\mu\nu}\nabla_{\mu}\nabla_{\nu}+V_2\right)\,, 
	\\
	\mathcal{O}_{3\pm}(\theta_0)&=M^{-1}\left(-g^{\mu\nu}D_{\mu}D_{\nu}+V_3\right)\,,
	\qquad
	D_{\mu}=\nabla_{\mu}\pm i\mathcal{A}_{\mu}\,, \nonumber 
	\\
	\mathcal{O}_{\pm}(\theta_0)&=M^{-\frac{1}{2}}\left(-i\left(\slashed{D}+\frac{1}{4}\slashed{\partial}\ln M\right)-i\Gamma_{\underline{01}}\left(1+V\right)\pm W\right)\,,
	\qquad
	D_{\mu}=\nabla_{\mu}\pm\frac{i}{2}\mathcal{A}_{\mu}\,, \nonumber 
\end{empheq}
with $g_{\mu\nu}$ and $\nabla_\mu$ evaluated for the $AdS_2$ metric \eqref{Eq:AdS2}, $\mathcal{A}_{\rho}=0$, $\mathcal{A}_{\tau}=\mathcal{A}$ and 
\begin{empheq}{alignat=7}\label{Eq:potential-a}
	V_2(\rho)&=-\frac{2\sin^2\theta(\rho)}{\sinh^2\rho}\,,
	\qquad
	V_3(\rho)=-\frac{\partial_{\rho}\mathcal{A}(\rho)}{\sinh\rho}\,,
	\qquad
	V(\rho)=\frac{1}{\sqrt{M(\rho)}}-1\,, 
	\\
	W(\rho)&=\frac{\sin^2\theta(\rho)}{\sqrt{M(\rho)}\sinh^2\rho}\,, \nonumber 
	\qquad
	\mathcal{A}(\rho)=1-\frac{1+\cosh\rho\cos\theta(\rho)}{\cosh\rho+\cos\theta(\rho)}\,. \label{Eq:potential-b}
\end{empheq}
The difference in 1-loop effective actions with the $\frac{1}{2}$-BPS string is then
\begin{empheq}{alignat=7}\label{eq: 1-loop effective action}
	e^{-\Delta\Gamma^{\textrm{1-loop}}_{\textrm{effective}}(\theta_0)}&=\left[\frac{\left(\frac{\det\,\mathcal{O}_+(\theta_0)}{\det\,\mathcal{O}_+(0)}\right)^{4}\left(\frac{\det\,\mathcal{O}_-(\theta_0)}{\det\,\mathcal{O}_-(0)}\right)^{4}}{\left(\frac{\det\,\mathcal{O}_1(\theta_0)}{\det\,\mathcal{O}_1(0)}\right)^{3}\left(\frac{\det\,\mathcal{O}_2(\theta_0)}{\det\,\mathcal{O}_2(0)}\right)^{3}\left(\frac{\det\,\mathcal{O}_{3+}(\theta_0)}{\det\,\mathcal{O}_{3+}(0)}\right)^{1}\left(\frac{\det\,\mathcal{O}_{3-}(\theta_0)}{\det\,\mathcal{O}_{3-}(0)}\right)^{1}}\right]^{\frac{1}{2}}\,.
\end{empheq}
The powers in the fermionic determinants reflect the Majorana nature of the spinors in Lorentzian signature.

The main difficulty in evaluating the above determinants is that the space is not homogeneous as is the case for $\theta_0=0$ where the results of \cite{Camporesi:1994ga,Camporesi:1995fb} are readily applied. A perturbative approach, valid for small values of $\theta_0$, was taken in \cite{Forini:2017whz} leading to the following evaluation of the one-loop effective action
\begin{empheq}{alignat=7}\label{eq: 1-loop effective action prediction}
	\Delta\Gamma^{\textrm{1-loop}}_{\textrm{effective}}(\theta_0)&=-\frac{3}{4}\theta_0^2+O\left(\theta_0^4\right)\,,
\end{empheq}
which coincides, to this order, with the expected field theory answer $\Delta\Gamma^{\textrm{1-loop}}_{\textrm{effective}}(\theta_0)=\frac{3}{2}\ln\cos\theta_0$ as follows from Eq. \ref{prediction}.  We will reproduce the perturbative result in this manuscript and consider the more general problem at arbitrary $\theta_0$.

%%%%%%%%%%%%%%%%%%%%%%%%%%%%%%%%%%%%%%%%%%%%%%%%%%%%
\section{Zeta-function regularization on $AdS_2$}\label{sec: results}

In this section we recall a number of results for determinants of Laplace- and Dirac-like operators in $AdS_2$ that  appear in a companion paper \cite{ZetaPaper} to which we refer the reader for the full derivation. The method  applies to operators defined on the $AdS_2$ geometry \eqref{Eq:AdS2}  and in the presence of external fields. Concretely, we consider general  operators of the form:
\begin{empheq}{alignat=7}\label{eq: 2d operator bosons}
	\mathcal{\bar O}&=-g^{\mu\nu}D_{\mu}D_{\nu}+m^2+V\,,
	&\qquad&\textrm{(bosons)}
	\\\label{eq: 2d operator fermions}
	\mathcal{\bar O}&=-i\left(\slashed{D}+\slashed{\partial}\Omega\right)-i\Gamma_{\underline{01}}\left(m+V\right)+W\,,
	&\qquad&\textrm{(fermions)}
\end{empheq}
with $D_{\mu}=\nabla_{\mu}-iq\mathcal{A}_{\mu}$. Under the assumption of circular symmetry, these operators can be expanded into their Fourier components
\begin{empheq}{alignat=7}
	\mathcal{\bar O}_l&=-\frac{1}{\sinh\rho}\partial_{\rho}\left(\sinh\rho\,\partial_{\rho}\right)+\frac{\left(l-q\mathcal{A}\right)^2}{\sinh^2\rho}+m^2+V\,,
	&\qquad
	l&\in\mathds{Z}\,,
	&\qquad&
	\textrm{(bosons)}
	\\
	\mathcal{\bar O}_l&=-i\Gamma_{\underline{1}}\left(\partial_{\rho}+\frac{1}{2}\coth\rho+\partial_{\rho}\Omega\right)+\Gamma_{\underline{0}}\frac{\left(l-q\mathcal{A}\right)}{\sinh\rho}-i\Gamma_{\underline{01}}\left(m+V\right)+W\,,
	&\qquad
	l&\in\mathds{Z}+\frac{1}{2}\,,
	&\qquad&
	\textrm{(fermions)}
\end{empheq}
where we have set $\mathcal{A}_{\rho}=0$ and $\mathcal{A}_{\tau}=\mathcal{A}(\rho)$, as well as $V=V(\rho)$, $W=W(\rho)$ and $\Omega=\Omega(\rho)$. Appropriate regularity conditions at the origin and fall-off conditions at infinity are required for  the background fields (see  \cite{ZetaPaper} for further details).

The ratio of determinants between the operators \eqref{eq: 2d operator bosons}-\eqref{eq: 2d operator fermions} and their free counterparts, obtained by setting $\mathcal{A}=\Omega=V=W=0$, is defined using $\zeta$-function regularization 
\begin{empheq}{alignat=7}\label{eq: renormalized determinant}
	\ln\frac{\det\mathcal{\bar O}}{\det\mathcal{\bar O}^{\textrm{free}}}&\equiv-\hat{\zeta}_{\mathcal{\bar O}}'(0)-\ln(\mu^2)\hat{\zeta}_{\mathcal{\bar O}}(0)\,,
	&\qquad
	\hat{\zeta}_{\mathcal{\bar O}}(s)&\equiv\zeta_{\mathcal{\bar O}}(s)-\zeta_{\textrm{free}}(s)\,,
\end{empheq}
where $\mu$ is the renormalization parameter. Extending previous results \cite{Dunne:2006ct} it was found in \cite{ZetaPaper} that these ratios are given by simple expressions.   The result for bosons reads
\begin{empheq}{alignat=7}
	\ln\frac{\det\mathcal{\bar O}}{\det\mathcal{\bar O}^{\textrm{free}}}&=\ln\frac{\det\mathcal{\bar O}_0}{\det\mathcal{\bar O}_0^{\textrm{free}}}+\sum_{l=1}^{\infty}\left(\ln\frac{\det\mathcal{\bar O}_l}{\det\mathcal{\bar O}_l^{\textrm{free}}}+\ln\frac{\det\mathcal{\bar O}_{-l}}{\det\mathcal{\bar O}_{-l}^{\textrm{free}}}+\frac{2}{l}\hat{\zeta}_{\mathcal{\bar O}}(0)\right)-2\left(\gamma+\ln\frac\mu2\right)\hat{\zeta}_{\mathcal{\bar O}}(0)\nn
	\\
	&+\int_0^{\infty}d\rho\,\sinh\rho\ln\left( \sinh\rho \right)V-q^2\int_0^{\infty}d\rho\,\frac{\mathcal{A}^2}{\sinh\rho}\,,\label{eq: main result bosons}
	\\
	\hat{\zeta}_{\mathcal{\bar O}}(0)&=-\frac{1}{2}\int_0^{\infty}d\rho\,\sinh\rho\,V\,,
	\label{z0bos}
\end{empheq}
whereas for fermions we have
\begin{empheq}{alignat=7}
	\ln\frac{\det\mathcal{\bar O}}{\det{\mathcal{\bar O}^{\textrm{free}}}}&=\sum_{l=\frac{1}{2}}^{\infty}\left(\ln\frac{\det\mathcal{\bar O}_l}{\det\mathcal{\bar O}_l^{\textrm{free}}}+\ln\frac{\det\mathcal{\bar O}_{-l}}{\det\mathcal{\bar O}_{-l}^{\textrm{free}}}+\frac{2}{l+\frac{1}{2}}\hat{\zeta}_{\mathcal{\bar O}}(0)\right)-2\left(\gamma+\ln\frac\mu2\right)\hat{\zeta}_{\mathcal{\bar O}}(0)\nn
	\\
	&+\int_0^{\infty}d\rho\,\sinh\rho\ln\left( \sinh\rho \right)\left(\left(m+V\right)^2-W^2-m^2\right)-q^2\int_0^{\infty}d\rho\,\frac{\mathcal{A}^2}{\sinh\rho}-\int_0^{\infty}d\rho\,\sinh\rho\,W^2,\label{24}
	\\
	\label{z0fer}
	\hat{\zeta}_{\mathcal{\bar O}}(0)&=-\frac{1}{2}\int_0^{\infty}d\rho\,\sinh\rho\left(\left(m+V\right)^2-W^2-m^2\right)\,,
\end{empheq}
with $\gamma\approx0.57721$,   the Euler-Mascheroni constant. In turn, the ratio for  Fourier modes is computed as
\begin{empheq}{alignat=7}\label{eq: ratio radial determinants}
	\frac{\det\mathcal{\bar O}_l}{\det\mathcal{\bar O}_l^{\textrm{free}}}&=\left\{
	\begin{array} {ll}
	{\displaystyle\frac{\psi_l(\infty)}{\psi_l^{\textrm{free}}(\infty)} }&,~~(\mathrm{bosons})\\
	\\
	{\displaystyle \frac{\psi_l^{(i)}(\infty)}{\psi_l^{(i)\,\textrm{free}}(\infty)}e^{\Omega(\infty)-\Omega(0)}}&,~~(\mathrm{fermions})
\end{array}\right.
\end{empheq}
%\begin{empheq}{alignat=7}\label{eq: ratio radial determinants}
%	\frac{\det\mathcal{O}_l}{\det\mathcal{O}_l^{\textrm{free}}}\frac{\det\mathcal{O}_{-l}}{\det\mathcal{O}_{-l}^{\textrm{free}}}&=\left\{
%	\begin{array} {ll}
%	{\displaystyle\frac{\psi_l(\infty)}{\psi_l^{\textrm{free}}(\infty)}\frac{\psi_{-l}(\infty)}{\psi_{-l}^{\textrm{free}}(\infty)} }&,~~(\mathrm{bosons})\\
%	\\
%	{\displaystyle \frac{\psi_l(\infty)}{\psi_l^{\textrm{free}}(\infty)}\frac{\psi_{-l}(\infty)}{\psi_{-l}^{\textrm{free}}(\infty)} e^{2(\Omega(\infty)-\Omega(0))}}&,~~(\mathrm{fermions})
%\end{array}\right.
%end{empheq}
where $\psi_l(\rho)$ is the solution to the homogeneous equation that is regular at $\rho=0$,
\begin{empheq}{alignat=7}
	\mathcal{\bar O}_l\psi_l&=0\,,
	&\qquad
	\psi_l(\rho)&\underset{\rho\rightarrow0}{\longrightarrow} \left\{ 
	\begin{array}{ll}
	{\displaystyle \rho^{|l|}}&, \,\quad (\mathrm{bosons})\\
	\\
	{\displaystyle \rho^{|l|-\frac{1}{2}}}&, \,\quad (\mathrm{fermions}).
	\end{array}\right.
\end{empheq}
For fermions  $\psi_l^{(i)}(\rho)$ is one (either) of the two components of the regular spinor solution to the \emph{first} order homogeneous equation. The overall normalization of $\psi_l$ in \eqref{eq: ratio radial determinants}  is not important as long as the leading coefficient of the small $\rho$ expansion matches that of the free solution\footnote{This is analogous to the usual  conditions $\psi(0)=0$, $\psi'(0)=1$ imposed on the homogeneous solutions in the application of the Gelfand-Yaglom method to 1d determinants  with Dirichlet boundary condition at the origin. In two and higher dimensions, the centrifugal barrier imposes the regular solution to vanish as a power law depending on the angular momentum, therefore generically $\psi'(0)=0$.}  $\psi_l^{\textrm{free}}$.

For the $\frac{1}{4}$-BPS strings we are interested in, the operators do not precisely take the form \eqref{eq: 2d operator bosons} or \eqref{eq: 2d operator fermions}, but rather they are conformally related to them
\begin{empheq}{alignat=7}\label{eq: rescaled operators bosons}
	\mathcal{O}&=M^{-1}\mathcal{\bar O}\,,
	&\qquad&
	\textrm{(bosons)}
	\\\label{eq: rescaled operators fermions}
	\mathcal{O}&=M^{-\frac{1}{2}}\mathcal{\bar O}\,.
	&\qquad&
	\textrm{(fermions)}
\end{empheq}
This can be understood as the effect of a Weyl rescaling of the metric by a function $M(\rho)$, which we assume to be smooth everywhere with $M(\rho)\rightarrow1$ for $\rho\rightarrow\infty$. Happily,  the determinants of $\mathcal{O} $ and $\mathcal{\bar O}$ are related by an anomaly calculation (cf. appendix A of \cite{Drukker:2000ep}). Indeed\footnote{ Boundary terms involving the extrinsic curvature and the normal derivative of the conformal factor do not contribute in the present case (see \cite{ZetaPaper} for details).},
\begin{empheq}{alignat=7}
	\ln\left(\frac{\det\mathcal{O}}{\det\mathcal{\bar O}^{\textrm{free}}}\right)&=\ln\left(\frac{\det\mathcal{\bar O}}{\det\mathcal{\bar O}^{\textrm{free}}}\right)+\frac{1}{4\pi}\int d^2\sigma\sqrt{g}\,\ln M\left[m^2+V-\frac{1}{6}R+\frac{1}{12}\nabla^2\ln M\right]
\label{33}
\end{empheq}
for bosons, while for fermions the result is
\begin{empheq}{alignat=7}
	\ln\left(\frac{\det\mathcal{O}}{\det\mathcal{\bar O}^{\textrm{free}}}\right)&=\ln\left(\frac{\det\mathcal{\bar O}}{\det\mathcal{\bar O}^{\textrm{free}}}\right)+\frac{1}{4\pi}\int d^2\sigma\sqrt{g}\,\ln M\left[\left(m+V\right)^2-W^2+\frac{1}{12}R-\frac{1}{24}\nabla^2\ln M\right]\,.
\label{34}
\end{empheq}

%%%%%%%%%%%%%%%%%%%%%%%%%%%%%%%%%%%%%%%%%%%
\section{One-loop effective action}\label{sec: Wilson loops}

In this section we apply the general results quoted in the previous section to the holographic description of the $\frac{1}{4}$-BPS latitude Wilson loops in $\mathcal{N}=4$ SYM \cite{Drukker:2007qr}. We refer the reader to the extensive literature for details; in particular to \cite{Forini:2015bgo,Faraggi:2016ekd,Forini:2017whz,Cagnazzo:2017sny}.

Before plunging into the calculation of each individual ratio in \eqref{eq: 1-loop effective action}, it is useful to combine the full spectrum of operators and gain some insight into the cancellations that occur in the one-loop effective action. Recall that, according to the discussion in section \ref{sec: results} (see eqns. \eqref{33} and \eqref{34}), the computation of each determinant is divided into two parts: an anomaly due to the Weyl transformation that maps the induced geometry \eqref{induced} to $AdS_2$, and the ratio for the corresponding rescaled operators. Notice that for the operators \eqref{eq: 1/4 BPS operators}, 
\begin{empheq}{alignat=7}
\label{frees}
{\cal O}(\theta_0=0)=\bar{\cal O}(\theta_0=0)=\bar {\cal O}^{\textrm{free}}.
\end{empheq}

Let us focus on the Weyl anomaly first. One can check that the potential and mass  terms  for the $\frac{1}{4}$-BPS operators \eqref{eq: 1/4 BPS operators} satisfy
\begin{empheq}{alignat=7}
\label{VV}
	8\times\left(\left(1+V\right)^2-W^2\right)-3\times2-3\times V_2-2\times V_3&=-R+\nabla^2\ln M\,,
\end{empheq}
a relation which is in fact a general feature of the gauge-fixed Nambu-Goto string, where the right hand side is recognized as the curvature of the induced metric, $R[M  g]=M^{-1}\left(R[g]-\nabla^2\ln M\right)$. The contribution from the curvature and conformal factor terms in \eqref{33}-\eqref{34} is
\begin{empheq}{alignat=7}
	\left(8\times\left(\frac{1}{12}\right)-8\times\left(-\frac{1}{6}\right)\right)R+\nabla^2\ln M\left(8\times\left(-\frac{1}{24}\right)-8\times\left(\frac{1}{12}\right)\right)&=2R-\nabla^2\ln M\,.
\end{empheq}
We then find that the modification to the ratio of determinants due to the rescaling of the metric is\footnote{The attentive reader may notice an unexpected non-trigonometric (linear) dependence $\theta_0$ in \eqref{ranomaly}. This comes about because the primitive involves inverse trigonometric functions which when evaluated at the endpoints and for $\theta_0\in(0,\frac\pi 2)$  simplify to the above expression. }
\begin{empheq}{alignat=7}
\label{ranomaly}
	\textrm{anomaly}:&
	&\qquad
	\frac{1}{4\pi}\int d^2\sigma\sqrt{g}\,R\ln M&=-\left(\theta_0\sin\theta_0+4\cos^2\frac{\theta_0}{2}\ln\cos\frac{\theta_0}{2}\right)\,.
\end{empheq}
Unlike the case where one maps the induced worldsheet metric to flat space \cite{Cagnazzo:2017sny}, the anomaly is non-vanishing\footnote{In that case the conformal factor is $M(\rho)\sinh^2\rho$. This becomes singular as $\rho\rightarrow\infty$, which forces the introduction of a large cut-off to regulate the divergences. Consequently, boundary terms must be added. These cancel, as does the bulk contribution since $R=0$.}. This is an effect of the curvature of $AdS_2$ and is perfectly compatible with the conformal invariance of the string action \cite{Drukker:2000ep} (see also  appendix B of \cite{Aguilera-Damia:2014bqa}).

We now move on to the computation of the determinants on $AdS_2$ using \eqref{eq: main result bosons} and \eqref{24}, starting with the total zeta-function  at the origin
\begin{empheq}{alignat=7}
\label{z0}
	\hat{\zeta}_{\textrm{tot}}(0)&=3\hat{\zeta}_{\mathcal{\bar O}_1}(0)+3\hat{\zeta}_{\mathcal{\bar O}_2}(0)+\hat{\zeta}_{\mathcal{\bar O}_{3+}}(0)+\hat{\zeta}_{\mathcal{\bar O}_{3-}}(0)-4\hat{\zeta}_{\mathcal{\bar O}_+}(0)-4\hat{\zeta}_{\mathcal{\bar O}_-}(0)\,.
\end{empheq}
This quantity determines the dependence of the one-loop effective action on the renormalization scale. Equations \eqref{z0bos} and \eqref{z0fer} show a slightly different combination of potentials than in \eqref{VV}, namely,
\begin{empheq}{alignat=7}
	8\times\left(\left(1+V\right)^2-W^2-1\right)-3\times V_2-2\times V_3&=\nabla^2\ln M\,,
\end{empheq}
which by itself does not vanish. When integrated, however,  it does,
\begin{empheq}{alignat=7}
	\int_0^{\infty}d\rho\,\sinh\rho\,\nabla^2\ln M&=\sinh\rho\,\partial_{\rho}\ln M\Big|_0^{\infty}=0
	&\qquad\Rightarrow\qquad
	\hat{\zeta}_{\textrm{tot}}(0)&=0\,.
\end{empheq}
As a consequence, no ambiguity related to the choice of renormalization scale, $\mu$, affects the effective action. The above cancellation also means that the Fourier sum of the combined bosons and fermions one-dimensional radial determinants does not need regularization\footnote{Each term in \eqref{z0} is responsible for  subtracting the divergence in the sum over Fourier modes in each individual determinant (see \eqref{eq: main result bosons} and \eqref{24}).}, in accordance with the calculations of \cite{Forini:2015bgo,Faraggi:2016ekd}.

A related quantity involving the same combination of potentials as $\hat{\zeta}_{\textrm{tot}}(0)$ is the sum of $\ln\left({\sinh\rho}\right)$ integrals in \eqref{eq: main result bosons} and \eqref{24}, which when added to the Weyl anomaly gives
\begin{empheq}{alignat=7}\label{eq: reminder cancelation}
	\textrm{anomaly}+\ln\sinh\rho:&
	&\qquad
	\int_0^{\infty}d\rho\,\sinh\rho\left(\frac{1}{2}R\ln M+\ln\left(\sinh\rho\right)\nabla^2\ln M\right)&=-2\ln\cos\frac{\theta_0}{2}\,.
\end{empheq}
As we will see, this terms cancels the reminder that was found in \cite{Forini:2015bgo,Faraggi:2016ekd}. We can also keep track of the contribution coming from the gauge field, easily seen to vanish:
\begin{empheq}{alignat=7}
	\mathcal{A}^2:&
	&\qquad
	1\times\left(1\right)^2+1\times\left(-1\right)^2-4\times\left(\frac{1}{2}\right)^2-4\times\left(-\frac{1}{2}\right)^2&=0\,.
\end{empheq}
In contrast, the last term in \eqref{24} involving the fermionic potential gives
\begin{empheq}{alignat=7}
	W^2:&
	&\qquad
	-8\times\int_0^{\infty}d\rho\,\sinh\rho\,W^2&=4\theta_0\sin\theta_0-16\sin^2\frac{\theta_0}{2}\,.
\end{empheq}
Ultimately, this expression accounts for the mismatch with the gauge theory prediction \eqref{prediction}.

Finally, one can check that the radial determinants at fixed Fourier mode coincide with those presented in \cite{Forini:2015bgo,Faraggi:2016ekd}. Therefore,
\begin{empheq}{alignat=7}\label{eq: sum over modes}
\sum\limits_{l}\ln \frac{\det\mathcal{\bar O}_l}{\det\mathcal{\bar O}_l^{\textrm{free}}}:&
	&\qquad
	-3\ln\cos\theta_0+2\ln\cos\frac{\theta_0}{2}\,.
\end{empheq}
In hindsight this was to be expected since the calculation involves solving a set of homogeneous equations in $AdS$ which translate into those of \cite{Forini:2015bgo,Faraggi:2016ekd} after an appropriate Weyl transformation of the metric and properly adjusting the potentials and connection terms. The difference in the present case is that instead of imposing a sharp Dirichlet boundary condition at small but finite  value of $\rho$ as in \cite{Forini:2015bgo,Faraggi:2016ekd}, here we only require regularity of the solutions at the center of the disk. Nevertheless, the answer is the same.  

Putting all the above results together, the final expression for the difference in the one-loop effective actions of the $\frac{1}{4}$ and $\frac{1}{2}$-BPS strings is
\begin{empheq}{alignat=7}
	\Delta\Gamma^{\textrm{1-loop}}_{\textrm{effective}}(\theta_0)&=\frac{3}{2}\ln\cos\theta_0+2\left(4\sin^2\frac{\theta_0}{2}-\theta_0\sin\theta_0\right)=-\frac{3}{4}\theta_0^2+O\left(\theta_0^4\right)\,.
\end{empheq}
As indicated above, when taking the small $\theta_0$ limit, our holographic answer coincides with the field theory prediction \eqref{prediction}, just as in the perturbative $\zeta$-function computation of \cite{Forini:2017whz}.

Let us briefly comment on this result. Recall that the works of \cite{Forini:2015bgo,Faraggi:2016ekd} computed the effective action by looking only at the sum of the radial determinants, finding the reminder $\ln\cos\frac{\theta_0}{2}$ in \eqref{eq: sum over modes}. Recently, it was argued in \cite{Cagnazzo:2017sny} that this term is corrected for if a diffeomorphism-invariant regulator is used in the calculation, producing a match between the string theory calculation and the gauge theory prediction. In contrast, the $\zeta$-function formalism is automatically diffeomorphism-invariant, and we see that this reminder disappears due to the combination \eqref{eq: reminder cancelation}. Alas, there is an extra contribution coming from the fermionic potential $W^2$ that yields a mismatch with the gauge theory calculation. At the moment we dare not speculate about the origin of this term.

For completness, we present the results for each individual determinant in the spectrum. Taking into account \eqref{frees}
\begin{empheq}{alignat=7}
	\ln\left(\frac{\det\,\mathcal{O}_1(\theta_0)}{\det\,\mathcal{O}_1(0)}\right)&=\theta_0\sin\theta_0+\frac{1}{2}\sin^2\frac{\theta_0}{2}+\left(\frac{7}{3}+2\cos\theta_0\right)\ln\left(\cos\frac{\theta_0}{2}\right)\nn
	\\
	&=\frac{7}{12}\theta_0^2+O\left(\theta_0^4\right)\,,
	\\
	\ln\left(\frac{\det\,\mathcal{O}_2(\theta_0)}{\det\,\mathcal{O}_2(0)}\right)&=-\theta_0\sin\theta_0+\frac{9}{2}\sin^2\frac{\theta_0}{2}+\left(\frac{7}{3}+2\cos\theta_0\right)\ln\left(\cos\frac{\theta_0}{2}\right) -2\ln\left(\Gamma\left(\cos\theta_0\right)\right)-\ln\left(\cos\theta_0\right)\nn
	\\ 
%	&~~~ -2\ln\left(\Gamma\left(\cos\theta_0\right)\right)-\ln\left(\cos\theta_0\right)\nn\\
	&=\left(\frac{1}{12}-\gamma\right)\theta_0^2+O\left(\theta_0^4\right)\,,
\\
	\ln\left(\frac{\det\,\mathcal{O}_{3\pm}(\theta_0)}{\det\,\mathcal{O}_{3\pm}(0)}\right)&=\frac{1}{2}\sin^2\frac{\theta_0}{2}+\left(\frac{7}{3}+2\cos\theta_0\right)\ln\left(\cos\frac{\theta_0}{2}\right)-\ln\left(\Gamma\left(\cos\theta_0\right)\right)-\ln\left(\cos\theta_0\right)\nn
	\\
	&=\frac{1}{2}\left(\frac{1}{6}-\gamma\right)\theta_0^2+O\left(\theta_0^4\right)\,,
\\
	\ln\left(\frac{\det\,\mathcal{O}_{\pm}(\theta_0)}{\det\,\mathcal{O}_{\pm}(0)}\right)&=\frac{1}{2}\theta_0\sin\theta_0+\left(\frac{7}{3}+2\cos\theta_0\right)\ln\left(\cos\frac{\theta_0}{2}\right)-\ln\left(\Gamma\left(\cos\theta_0\right)\right)-\ln\left(\cos\theta_0\right)\nn
	\\
	&=\frac{1}{2}\left(\frac{11}{12}-\gamma\right)\theta_0^2+O\left(\theta_0^4\right)\,.
\end{empheq}
Our results match the perturbative heat kernel calculation of \cite{Forini:2017whz}. Notice that the first ratio is entirely an effect of the Weyl anomaly, since the rescaled operators for the $\frac{1}{4}$-BPS and the $\frac{1}{2}$-BPS solutions coincide. Actually, we have checked that all the ratios for the rescaled operators, without including the anomaly, also match with the perturbative method for a fixed $AdS_2$ metric. It would be interesting to extend the perturbative heat kernel results of \cite{Forini:2017whz} to the next order in $\theta_0$.

%%%%%%%%%%%%%%%%%%%%%%%%%%%%%%%%%%%%%%%%%%%%%%%%%%%%%
\section{Conclusions}\label{Sec:Conclusions}

In this manuscript we have computed the difference of one-loop effective actions of the $\frac {1}{4}$-  and  $\frac{1}{2}$-BPS strings using $\zeta$-function regularization. We were encouraged and motivated by a previous perturbative heat kernel computation reporting agreement with the field theory prediction at the  first nontrivial order in the latitude angle $\theta_0$ \cite{Forini:2017whz}. It is worth highlighting that we tackled the computation directly on the hyperbolic disk rather than mapping the problem to a cylinder, as has been traditionally done  \cite{Kruczenski:2008zk,Forini:2015bgo,Faraggi:2016ekd,Cagnazzo:2017sny}.  Along these lines, it would be an elucidating step to adapt our results to compute the $\zeta$-function for circularly symmetric operators defined on the flat cylinder geometry. This would shed some light on the role of the diffeomorpism-invariant regulator advocated in \cite{Cagnazzo:2017sny}. We hope to pursue these directions in the near future.

 Alas, our complete computation shows that at higher order in $\theta_0$ the agreement is lost. We are thus, left facing a puzzle. Armed with the supersymmetric localization answer we can indulge in a form of answer analysis. As stated before, the remainder of previous calculations does not appear in our approach since $\zeta$-function regularization is explicitly diffeormorphism invariant.  One identifiable culprit for the discrepancy we now faced is the term proportional to $W^2$ in the expression for the fermions. We suspect that ultimately some aspects of chiral symmetry might be at play, as suggested in \cite{Avan:1984bq} in a different context.  Another potential problem underlying our discrepancy could be supersymmetry. We do not see how to move forward in this direction at the moment but find it quit plaussible to be the cause of the discrepancy. 

Our manuscript is a push in understanding the role of  technical methods needed to tackle precision computations in holography and we are certain that its application will go beyond the one presented here. We hope to return, for example, to a similar computation in the context of the ABJM duality.  It is also plausible that the methods systematically developed in our companion paper \cite{ ZetaPaper} and used explicitly here, will find use in other problems possibly related to one-loop supergravity computations in the context of corrections to the black hole entropy.

\section*{Acknowledgments}

We are grateful to F. Larsen, D Trancanelli, A. Tseytlin and E. Vescovi.
We are particularly thankful to K. Zarembo for various pointed comments. LPZ, VR and GAS thank ICTP for providing  hospitality at various stages. 
AF was supported by Fondecyt \# 1160282.  LPZ and VR are partially supported by the US Department of Energy under Grant No. DE-SC0017808 –{\it  Topics in the AdS/CFT Correspondence: Precision tests with Wilson loops, quantum black holes and dualities.} GAS and JAD are supported by CONICET and grants PICT
2012-0417, PIP0595/13, X648 UNLP, PIP 0681 and PI {\it B\'usqueda de nueva F\'isica}.

\bibliographystyle{JHEP}
\bibliography{Bib}
\end{document}